# Performance Analysis of Millimeter-Wave Relaying: Impacts of Beamwidth and Self-Interference

Guang Yang, and Ming Xiao, *Senior Member, IEEE*

*Abstract*—We study the maximum achievable rate of a two-hop amplified-and-forward (AF) relaying millimeter-wave (mm-wave) system, where two AF relaying schemes, i.e., half-duplex (HD) and full-duplex (FD) are discussed. By considering the two-ray mm-wave channel and the Gaussian-type directional antenna, jointly, the impacts of the beamwidth and the self-interference coefficient on maximum achievable rates are investigated. Results show that, under a sum-power constraint, the rate of FD-AF mm-wave relaying outperforms its HD counterpart only when antennas with narrower beamwidth and smaller self-interference coefficient are applied. However, when the sum-power budget is sufficiently high or the beamwidth of directional antenna is sufficiently small, direct transmission becomes the best strategy, rather than the AF relaying schemes. For both relaying schemes, we show that the rates of both AF relaying schemes scale as $\mathcal{O}\left(\min\left\{\theta_m^{-1}, \theta_m^{-2}\right\}\right)$ with respect to beamwidth $\theta_m$, and the rate of FD-AF relaying scales as $\mathcal{O}\left(\mu^{-\frac{1}{2}}\right)$ with respect to self-interference coefficient $\mu$. Besides, we show that, ground reflections may significantly affect the performance of mm-wave communications, constructively or destructively. Thus, the impact of ground reflections deserves careful considerations for analyzing or designing future mm-wave wireless networks.

## I. INTRODUCTION

Thanks to abundant spectrum resources, wireless communications at millimeter-wave (mm-wave) bands (ranging from around 24 GHz to 300 GHz) is a key enabler for fulfilling the multi-Gbps rates in future mobile communications [1], [2]. However, mm-wave communications still need to overcome many challenges. A typical challenge is the severe path loss. One common solution is to use highly directional antennas. Normally, a directional antenna with narrower beamwidth has a higher antenna gain at the main lobe, which can compensate for the path loss, and hence can significantly improve the transmission distance and reduce the outage probability [3]–[6]. In previous works, the radiation pattern of directional antennas was usually modeled in an idealized fashion, i.e., a large constant antenna gain within the narrow main-lobe and zero elsewhere. This idealized radiation pattern (often referred to as the "flat-top" model) was widely used [7]–[9] for performance analysis, due to its high tractability. Evidently, the side-lobe effect is neglected in the flat-top model, which however is not negligible when the beamwidth is less than 60°, as demonstrated in [10]. In recent years, numerous efforts have been dedicated in directional antennas, and the effect of side-lobe leakages is incorporated. For instance, in [11], [12], a piece-wise model was applied. In [6], [13]–[15], a sectorized model that considers the side-lobe leakage is employed, which has both constant gains in the main and side lobes, respectively.

In addition to the severe path loss, mm-wave communications suffer high penetration loss and weak diffusion, diffraction, and high-order reflections [16]. Therefore, mm-wave radios are more suitable for short-range wireless systems. To achieve the higher coverage and robustness, relaying techniques may be used for mm-wave communications. According to the duplex mode, relaying techniques can be categorized as half-duplex (HD) and full-duplex (FD). In contrast to the HD relay, the FD relay can support simultaneous reception and transmission, while it suffers from the self-interference. Numerous efforts have been devoted to exploring potentials of performance enhancement via relays in mm-wave communications [17]–[20]. In [21], the energy efficiency was studied for 60 GHz indoor networks with amplify-and-forward (AF) relay. The comparison between two duplex modes was performed, and the impacts of imperfect self-interference suppression, drain efficiency and static circuit power were also studied. In recent works [22], [23], the algorithm of energy-efficient cross-layer resource allocation with for the FD decode-and-forward (DF) relaying was developed.

However, for mm-wave relaying systems with directional antennas, there are two major limitations in the existing works:

(i) The sectorized model (e.g., [6], [14], [15]) is a prevailing option when modeling the radiation pattern of directional antennas, due to its high tractability for analysis. However, in the sectorized model, only two constant gains are used to characterize the main and side lobes, respectively, without any transition between them. An obvious drawback of this idealization is that, the crucial "roll-off" feature (a gradual decay from the main lobe to the side lobe) of the real-world radiation pattern for directional antennas is not reflected, and the resulting discontinuity may seriously affect the system performance evaluations.

(ii) It has been shown in [24]–[27] that, first-order reflections are not negligible in mm-wave communications. However, in most of preceding works (e.g., [14], [16], [28], [29]), the impacts of ground reflections (first-order reflections) are rarely incorporated, since it is widely and deeply believed that the ground reflection is not a dominant factor that can dramatically affect performance

This work was supported by EU Marie Curie Project, QUICK, No. 612652, and Wireless@KTH Seed Project "Millimeter Wave for Ultra-Reliable Low-Latency Communications".

Guang Yang and Ming Xiao are with Department of Information Science and Engineering (ISE), KTH Royal Institute of Technology, Stockholm, Sweden (Email: {gy, mingx}@kth.se).

evaluations. This conventional channel model for mm-wave radios (based on the LOS path only) may lead to an obvious overestimation (resp., underestimation) in performance evaluations, due to the omission of the non-negligibly constructive (resp., destructive) effects of reflections.

It is evident that, the aforementioned limitations may result in obvious inaccuracy when analyzing the performance of mm-wave wireless communications. To address above two problems, two heuristic models are considered in this paper. For the radiation pattern of directional antennas, the *Gaussian-type directional antenna model* [30]–[33] is used, where the main-lobe gain attenuates to a non-zero side-lobe gain in the continuous manner, such that the "roll-off" feature of the real radiation pattern can be seized, while preserving the tractability. Furthermore, to incorporate ground reflections in mm-wave channels, the *two-ray channel model* [34]–[36] is employed. Due to the fact that ground normally acts as the commonest reflective surface in various scenarios[1], we consider ground reflections only in our study. The use of the two-ray model in mm-wave communications is fairly recent, which can be found in [37]–[40].

The objective of our paper is to study the maximum achievable rates of a two-hop AF relaying system with mm-wave. More precisely, we first formulate the rates by HD-AF and FD-AF relaying schemes, respectively, where the optimized time-sharing scheme for the HD mode and optimal power allocations are studied. To the best of our knowledge, we are the first to consider a joint treatment of antenna model and channel model, which is important for analyzing the performance of mm-wave communications. Subsequently, we investigate the impacts of the beamwidth and the self-interference cancellation on mm-wave communications with AF relays. The main contributions of our work can be summarized from the following aspects:

- To overcome the limitations induced by the conventional oversimplification of the directional antenna and the propagation model for mm-wave communications, we consider the Gaussian antenna model and the two-ray channel model, *jointly*[2]. Note that, there exists a trade-off between the main-lobe and side-lobe gains in the Gaussian antenna model when varying the beamwidth. Paired with the two-ray channel model, evidently, the distribution of signal strength on direct and reflected propagation paths heavily depends on the beamwidth. Thus, a joint treatment of two heuristic models is essential, and it enables performance evaluations of mm-wave systems in a more accurate way.
- In terms of main-lobe beamwidth $\theta_m$, we demonstrate that the maximum rates of AF relaying schemes scale as $\mathcal{O}\left(\min\left\{\theta_m^{-1}, \theta_m^{-2}\right\}\right)$. Furthermore, let self-interference coefficient $\mu \in (0, 1)$ characterize the self-interference cancellation. That is, a smaller $\mu$ indicates better self-interference cancellation, we find that the rate of FD-AF relaying scales as $\mathcal{O}\left(\mu^{-\frac{1}{2}}\right)$. The convexity of the decay in the achievable rate with respect to $\mu$ reveals great benefits by strengthening the self-interference cancellation in FD-AF relay implementations. These results give a more convenient way to keep track of the performance, and provide insights for the future system design.
- In contrast to the conventional belief that the effect of reflections is negligible in mm-wave communications, our results show that, the constructive or destructive contribution of ground reflections is indeed nontrivial. Note that, the ground reflection heavily relies on the incident angle, the radiation pattern of directional antenna, and the transmission distance, the contribution of ground reflections may be fairly obvious. Thus, careful considerations regarding the impacts of ground reflections are valuable for analyzing or designing mm-wave networks.

The rest of the paper is organized as follows. In Sec. II, we give a system for two-hop AF relaying with mm-wave, where the two-ray channel and Gaussian antennas are jointly considered. In Sec. III, the maximum achievable rates of two-hop relaying mm-wave systems by two relaying schemes are presented, respectively, where time-sharing and power allocations are incorporated. In Sec. IV, the impacts of the beamwidth and the self-interference cancellation on the rates of two relaying schemes are comprehensively studied. Numerical results are provided in Sec. V, which are followed by conclusions drawn in Sec. VI.

In our paper, we use "big-$\mathcal{O}$" notation to denote the variation of rates with respect to the beamwidth or the self-interference coefficient. The big-$\mathcal{O}$ notation is defined as: let $u(x)$ and $f(x)$ be two functions defined on some subset of the real numbers, then we let $u(x) = \mathcal{O}(f(x))$ whenever $|u(x)/f(x)|$ is upper bounded for all feasible $x$.

## II. SYSTEM MODEL

We consider a two-hop AF relaying system with mm-wave, which consists of a source node $S$, a destination node $D$, and a relay node $R$ (in the HD or FD mode)[3], as illustrated in Fig. 1. All nodes are equipped with directional antennas. For notational simplicity, we denote by $h_i \in \mathbb{C}$, $i \in \{1, 2\}$ the channel coefficients of links $S - R$ and $R - D$, respectively. We denote by $H_S$, $H_R$ and $H_D$ the deployment height of $S$, $R$ and $D$, respectively. In addition, for $S - R$ and $R - D$ links, we let $L_1$ and $L_2$ be the respective horizontal separation distances.

In this section, we elaborately show the two-ray channel model and the Gaussian antenna model, respectively, which are used for our subsequent performance analysis for mm-wave communications. At mm-wave bands, the multi-path effect resulted by scattering is negligible, since electromagnetic waves with short wavelength have weak capability of diffraction or high-order reflections [16]. Also, the shadowing effect happens only when obstacles emerge in the propagation path of mm-wave radios, and severe shadowing may result in the link blockage. To mitigate the detrimental effect by shadowing, one can elevate the deployment heights of nodes, i.e., $H_S$, $H_R$, and $H_D$, such that the probability that obstacles appear in wave propagation path can be reduced. For focusing on impacts of the beamwidth and self-interference, both multipath and shadowing effects are not considered for mm-wave channels in our study.

---

[1]The channel model, of course, can be generalized to a "multi-ray" version, if multiple dominant reflective objects simultaneously exist.

[2]It is worth mentioning that, the Gaussian antenna model itself is not novel. However, the joint treatment of Gaussian antenna model and two-ray channel model has not been performed previously, which is one of our main contributions.

[3]We assume that there exists no direct link between $S$ and $D$ in our scenario (e.g., due to extremely severe blockage or path loss).

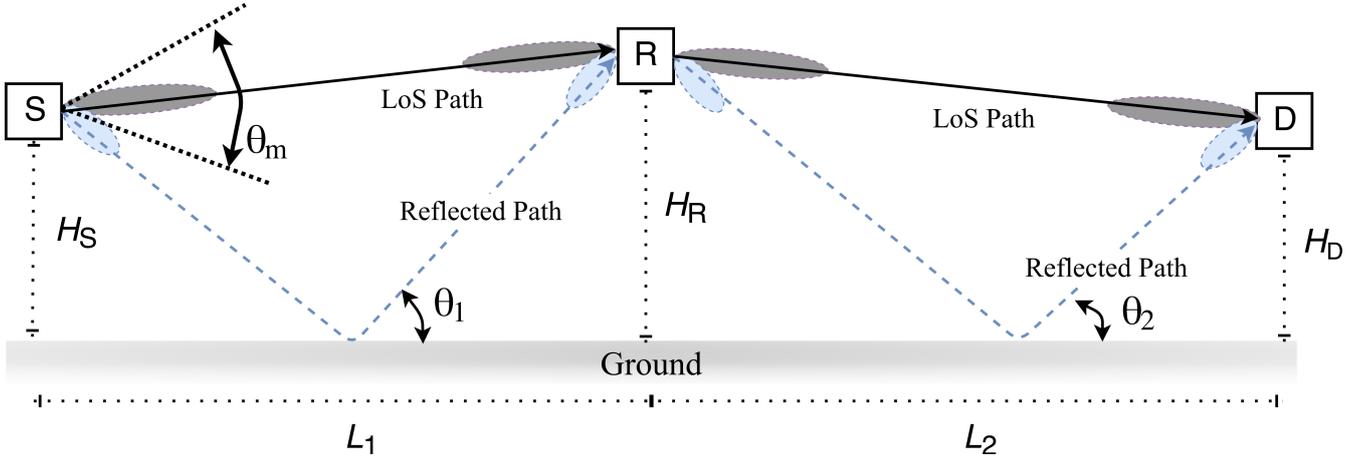

Fig. 1. Two-hop AF relaying system: two-ray channel and directional antennas.

*A. Two-Ray Channel*

The two-ray model is a classic channel model that considers two major coexisting transmission paths, i.e., the light-of-sight (LOS) path and the reflection path, and the feasibility of modeling mm-wave channels has been investigated in [36]–[38]. The two-ray channel model is shown in Fig. 1.

By Friis transmission formula, the received signal power is written as $P_r = P_t \cdot |h_i|^2$, where $P_t$ denotes the transmit power, and $h_i$ for $i = 1, 2$ denotes the channel coefficient. In the presence of reflections from the ground surface, $h_i$ is exactly written as

$$\begin{cases} h_1 = \dfrac{\lambda \left(G(0) + G(\theta_1)\,\Gamma(\theta_1)\cos(\theta_1)\,e^{-j\Delta\varphi_1}\right)}{4\pi\sqrt{(H_S - H_R)^2 + L_1^2}} \\ h_2 = \dfrac{\lambda \left(G(0) + G(\theta_2)\,\Gamma(\theta_2)\cos(\theta_2)\,e^{-j\Delta\varphi_2}\right)}{4\pi\sqrt{(H_R - H_D)^2 + L_2^2}} \end{cases}, \quad (1)$$

where $\lambda$ is the wavelength of mm-wave radio, $G(*)$ denotes the radiation pattern of directional antenna (to be elaborated subsequently), and $\theta_i$ denotes the reflection angle relative to the ground plane, which is given as

$$\begin{cases} \theta_1 = \arctan\left(\dfrac{H_S + H_R}{L_1}\right) \\ \theta_2 = \arctan\left(\dfrac{H_R + H_D}{L_2}\right) \end{cases}.$$

Furthermore, the phase difference $\Delta\varphi_i$ for $i = 1, 2$, which is characterized by the length difference between the LOS path and the reflection path, is expressed as

$$\begin{cases} \Delta\varphi_1 = \dfrac{2\pi}{\lambda}\left(\sqrt{(H_S + H_R)^2 + L_1^2} - \sqrt{(H_S - H_R)^2 + L_1^2}\right) \\ \Delta\varphi_2 = \dfrac{2\pi}{\lambda}\left(\sqrt{(H_R + H_D)^2 + L_2^2} - \sqrt{(H_R - H_D)^2 + L_1^2}\right) \end{cases}.$$

In addition, reflection coefficient $\Gamma(\theta_i)$ associated with reflection angle $\theta_i$ [35] is given as

$$\Gamma(\theta_i) = \frac{\sin\theta_i - Z(\theta_i)}{\sin\theta_i + Z(\theta_i)}.$$

Here, $Z(\theta_i)$ with respect to perpendicularly and horizontally polarized electromagnetic waves are respectively shown as

$$Z(\theta_i) = \begin{cases} \omega^{-1}\sqrt{\omega - \cos^2(\theta_i)}, & \text{perp. polarization} \\ \sqrt{\omega - \cos^2(\theta_i)}, & \text{horiz. polarization} \end{cases}, \quad (2)$$

where $\omega$ denotes the dielectric constant of ground.

*B. Gaussian-Type Directional Antenna*

For radiation pattern $G(*)$, we consider a Gaussian-type directional antenna model, which captures the "roll-off" feature of real radiation patterns [31]–[33], and matches well with measured results [30]. For the validation of the Gaussian-type directional antenna model, interested readers are referred to related literature, e.g., [30]–[33].

In the Gaussian-type model[4], let $\phi \in [-\pi, \pi)$ be the orientation angle relative to the boresight. With respect to the constraint for the total radiated power in all directions, i.e., $\int_{-\pi}^{\pi} G(\theta)d\theta = 2\pi$, the antenna gain along this orientation is given as

$$G(\phi) = \frac{2\pi}{V(\theta_m, \theta_h) + 2\pi - \theta_m} 10^{\frac{3}{10}\left[\frac{\theta_m^2 - 4\phi^2}{\theta_h^2}\right]_+}, \quad (3)$$

---

[4]Slightly different from the model adopted in [30]–[33], we in this paper particular introduce the constraint of total radiated power in all directions. Details are provided in the subsequent section.

where $[*]_+ \triangleq \max\{*, 0\}$, $\theta_h$ is the half-power beamwidth, $\theta_m$ is the main-lobe beamwidth, and $V(\theta_m, \theta_h)$ is defined as

$$V(\theta_m, \theta_h) \triangleq \int_0^{\theta_m} 10^{\frac{3}{10}\left(\frac{\theta_m^2 - x^2}{\theta_h^2}\right)} dx.$$

According to evaluations performed in [30], [32], we have the empirical expression $\theta_m = 2.6 \cdot \theta_h$ for $\frac{\pi}{12} \leq \theta_m \leq \frac{\pi}{3}$. Thus, (3) can be further reduced to

$$G(\phi) = \frac{2\pi}{2\pi + 42.6443\theta_m} \cdot 10^{2.028\left[1 - \left(\frac{2\phi}{\theta_m}\right)^2\right]_+}, \quad (4)$$

which will be used throughout the paper.

## III. RATES OF AF RELAYING SCHEMES

In this section, we study the maximum achievable rates of two-hop AF relaying system with mm-wave, as illustrated in Fig. 1. The channel gain of a two-ray mm-wave channel on the $i^{\text{th}}$ hop for $i = 1, 2$ is characterized by $g_i \triangleq |h_i|^2$, following (1), and the antenna gain of Gaussian-type directional antenna follows the expression in (4).

For notational simplicity, we define $\xi_i$ as the transmit power on the $i^{\text{th}}$ hop. Besides, let $s$ be the signal-to-noise ratio (SNR) or the signal-to-interference-plus-noise ratio (SINR), then we use $C(s) \triangleq \log_2(1+s)$ to denote the achievable rate. In our study, sum-power constraint is considered for both AF relaying schemes[5], which will be elaborated subsequently.

### A. Half-Duplex Mode (HD-AF)

Let $x$ be the signal transmitted from $S$, where $x \sim \mathcal{CN}(0, 1)$ is independent and identically distributed (i.i.d.). Then, the received signal at $R$ is given as

$$y_R = \sqrt{\xi_1} h_1 \cdot x + n_1,$$

where the AWGN $n_1$ at $R$ follows $\mathcal{N}(0, 1)$. After amplifying $y_R$ by scaling coefficient $a$, i.e.,

$$a = \sqrt{\frac{\xi_2}{g_1 \xi_1 + 1}}.$$

Then, the amplified signal is subsequently sent to $D$ from $R$, and the signal that finally reaches $D$ is obtained as

$$y_D = a\sqrt{\xi_1} h_1 h_2 \cdot x + a h_2 \cdot n_1 + n_2,$$

where the AWGN $n_2$ at $D$ follows $\mathcal{N}(0, 1)$.

Given a sum-power constraint, power allocation is performed to maximize the rate of HD-AF relaying. We consider the optimized time-sharing strategy for HD-AF relaying. A normalized time-sharing parameter $0 \leq \beta \leq 1$ is used to characterize the duration for the transmission from $S$ to $R$, and $(1 - \beta)$ is left for the transmission from $R$ to $D$. Thus, the sum-power constraint that incorporates the time-sharing scheme can be interpreted as

$$\beta \xi_1 + (1 - \beta) \xi_2 = \xi, \quad (5)$$

[5]Sum-power constraint can be realized by a centralized power controller, and it is already used in many works on optimizing multi-hop networks. e.g., [41], [42]. The study also can be extended to the scenario with per-node power constraint, which however is not considered in our study, since our main focus is to investigate the impacts of beamwidth and self-interference.

where $\xi$ represents the sum-power constraint in terms of $\xi_1$ and $\xi_2$. In the following proposition, we formulate the maximum achievable rate of HD-AF by applying time sharing and power allocation.

**Proposition 1.** *For two-hop HD-AF relaying, given sum-power constraint $\beta \xi_1 + (1 - \beta) \xi_2 = \xi$, the maximum achievable rate is formulated as*

$$\eta_{\text{HD}}^* = \max_{\substack{0 \leq \beta \leq 1 \\ \xi_1, \xi_2 \geq 0 \\ \beta \xi_1 + (1-\beta)\xi_2 = \xi}} \min \left\{ \begin{array}{c} \beta C(g_1 \xi_1), \\ (1-\beta) C\left(\frac{g_1 g_2 \xi_1 \xi_2}{1 + g_1 \xi_1 + g_2 \xi_2}\right) \end{array} \right\}.$$

*Proof:* From $S$ to $R$, it is clear that, the throughput with respect to duration $\beta$ is given as

$$T_{S-R} = \beta \log_2(1 + g_1 \xi_1) \triangleq \beta C(g_1 \xi_1).$$

From $R$ to $D$, due to the AF policy, the transmitted signal from relay is a scaled version of its received signal, and hence the throughput is obtained as

$$T_{R-D} = (1-\beta) C\left(\frac{g_1 g_2 \xi_1 \xi_2}{1 + g_1 \xi_1 + g_2 \xi_2}\right).$$

Since the end-to-end throughput is restricted to the bottleneck hop, the overall throughput is $T = \min\{T_{S-R}, T_{R-D}\}$. With (5), the optimization problem can be formulated. ∎

However, it is worth noting that, it is a non-trivial task to solve the optimization problem formulated in Proposition 1. The major difficulty lies in the constraint introduced by (5), where the selection of the optimal time-sharing parameter and power allocation are intertwined with each other. More precisely, the optimal $\beta$ depends on $\xi_1$ and $\xi_2$, while $\xi_1$ and $\xi_2$ are constrained to $\xi$ in terms of $\beta$ via (5). In this case, we are unable to decouple the constraints straightforwardly, such that it is intractable to obtain a closed-form solution for $\eta_{\text{HD}}^*$. However, in what follows, we will study the maximum achievable rate in a special case, i.e., with fixed $\beta = \frac{1}{2}$, which corresponds to HD-AF relaying with equally assigned time slots for two transmission phases. Thus, with fixed $\beta = \frac{1}{2}$, the optimization problem in Proposition 1 can be reduced to

$$\eta_{\text{HD}}^* \bigg|_{\beta = \frac{1}{2}} = \max_{\substack{\xi_1 + \xi_2 = 2\xi \\ \xi_1, \xi_2 \geq 0}} \frac{1}{2} C\left(\frac{g_1 g_2 \xi_1 \xi_2}{1 + g_1 \xi_1 + g_2 \xi_2}\right)$$

$$\triangleq \max_{\substack{\xi_1 + \xi_2 = 2\xi \\ \xi_1, \xi_2 \geq 0}} \frac{1}{2} C(\gamma_{\text{HD}}(\xi_1, \xi_2)) = \frac{1}{2} C\left(\max_{\substack{\xi_1 + \xi_2 = 2\xi \\ \xi_1, \xi_2 \geq 0}} \gamma_{\text{HD}}(\xi_1, \xi_2)\right).$$

For the maximum SNR, denoted by $\gamma_{\text{HD}}^*$, we have

$$\gamma_{\text{HD}}^* \triangleq \max_{\substack{\xi_1 + \xi_2 = 2\xi \\ \xi_1, \xi_2 \geq 0}} \gamma_{\text{HD}}(\xi_1, \xi_2) = \max_{\substack{\xi_1 + \xi_2 = 2\xi \\ \xi_1, \xi_2 \geq 0}} \frac{g_1 g_2 \xi_1 \xi_2}{1 + g_1 \xi_1 + g_2 \xi_2}$$

$$= \left(\min_{0 \leq \xi_1 \leq 2\xi} \frac{1 + (2g_2 \xi)^{-1}}{g_1 \xi_1} + \frac{1 + (2g_1 \xi)^{-1}}{g_2(2\xi - \xi_1)}\right)^{-1}$$

$$= \frac{4 g_1 g_2 \xi^2}{\left(\sqrt{1 + 2g_1 \xi} + \sqrt{1 + 2g_2 \xi}\right)^2},$$

which concludes the maximum spectral efficiency, i.e.,

$$\eta_{\text{HD}}^*\Big|_{\beta=\frac{1}{2}} = \frac{1}{2}C\left(\frac{4g_1g_2\xi^2}{\left(\sqrt{1+2g_1\xi}+\sqrt{1+2g_2\xi}\right)^2}\right), \quad (6)$$

whenever the power allocation below can be applied:

$$\begin{cases} \xi_1^* = \dfrac{2\xi\sqrt{1+2g_2\xi}}{\sqrt{1+2g_1\xi}+\sqrt{1+2g_2\xi}} \\ \xi_2^* = \dfrac{2\xi\sqrt{1+2g_1\xi}}{\sqrt{1+2g_1\xi}+\sqrt{1+2g_2\xi}} \end{cases}.$$

## B. Full-Duplex Mode (FD-AF)

In FD-AF relaying, $\mu \in (0,1)$ is the self-interference coefficient, which characterizes the capability of self-interference techniques[6]. That is, a smaller $\mu$ indicates a more powerful self-interference suppression. Let $x[k]$ denote the transmitted signal from $S$ at time slot $k$, and we assume it is i.i.d. over time slots. At $R$, it is easy to obtain that

$$\begin{cases} y_R^{(r)}[k] = \sqrt{\xi_1}h_1 \cdot x[k] + \sqrt{\mu}y_R^{(t)}[k-1] + n_1[k] \\ y_R^{(t)}[k] = a \cdot y_R^{(r)}[k] \end{cases},$$

where $y_R^{(r)}[k]$ and $y_R^{(t)}[k]$ represent the received and transmitted signals at $R$ in time slot $k$, respectively, the variance of AWGN $n_1[k]$ is 1, and the scaling coefficient $a$ is given as

$$a = \sqrt{\frac{\xi_2}{\xi_1 g_1 + 1 + \mu \xi_2}}.$$

Finally, the received signal at $D$ can be expressed as

$$y_D[k] = ah_2\left(\sqrt{\xi_1}h_1 x[k] + \sqrt{\mu}y_R^{(t)}[k-1] + n_1[k]\right) + n_2[k],$$

where the variance of AWGN $n_2[k]$ is 1. Thus, the SINR can be easily obtained as

$$\gamma_{\text{FD}}(\xi_1, \xi_2) = \frac{g_1 g_2 \xi_1 \xi_2}{(g_2\xi_2 + 1)(\mu\xi_2 + 1) + g_1\xi_1}, \quad (7)$$

and the resulting rate is given by $\eta_{\text{FD}} = C(\gamma_{\text{FD}}(\xi_1, \xi_2))$.

Given a sum-power constraint, in following Proposition 2, the maximum achievable rate of FD-AF relaying by power allocation is presented, where the sum-power constraint is given as $\xi_1 + \xi_2 = \xi$.

**Proposition 2.** *For two-hop FD-AF relaying, given the sum-power constraint $\xi_1 + \xi_2 = \xi$, the maximum achievable rate is*

$$\eta_{\text{FD}}^* = C\left(\frac{g_1 g_2 \xi^2}{2+(g_1+g_2+\mu)\xi+2\sqrt{(1+g_1\xi)(1+g_2\xi)(1+\mu\xi)}}\right),$$

*where the optimal power allocation is given as*

$$\begin{cases} \xi_1^* = \dfrac{\xi\sqrt{1+(\mu+g_2)\xi+\mu g_2\xi^2}}{\sqrt{1+g_1\xi}+\sqrt{1+(\mu+g_2)\xi+\mu g_2\xi^2}} \\ \xi_2^* = \dfrac{\xi\sqrt{1+\xi g_1}}{\sqrt{1+\xi g_1}+\sqrt{1+(\mu+g_2)\xi+\mu g_2\xi^2}} \end{cases}.$$

[6]Commonly, the self-interference coefficient is determined by the back-end implementation, circuit design or signal processing techniques, which is independent of the beamwidth of directional antennas at the front-end.

*Proof:* We know that $\eta_{\text{FD}}^* = C(\gamma_{\text{FD}}^*)$, where $\gamma_{\text{FD}}^*$ denotes the maximized SINR at destination, i.e.,

$$\begin{aligned}\gamma_{\text{FD}}^* &\triangleq \max_{\substack{\xi_1+\xi_2=\xi \\ \xi_1,\xi_2\geq 0}} \gamma_{\text{FD}}(\xi_1,\xi_2) \\ &= \max_{\substack{\xi_1+\xi_2=\xi \\ \xi_1,\xi_2\geq 0}} \frac{g_1g_2\xi_1\xi_2}{(g_2\xi_2+1)(1+\mu\xi_2)+g_1\xi_1} \\ &= \left(\min_{0\leq \xi_1\leq \xi} \frac{1+\mu\xi+\frac{\mu+\frac{1}{\xi}}{g_2}}{g_1\xi_1}+\frac{1+\frac{1}{g_1\mu}}{g_2\xi_2}-\frac{\mu}{g_1}\right)^{-1} \\ &= \frac{g_1g_2\xi^2}{\left(\sqrt{1+g_1\xi}+\sqrt{1+(\mu+g_2)\xi+\mu g_2\xi^2}\right)^2 - \mu g_2\xi^2} \end{aligned} \quad (8)$$

whenever the power allocation below can be applied:

$$\begin{cases} \xi_1^* = \dfrac{\xi\sqrt{1+(\mu+g_2)\xi+\mu g_2\xi^2}}{\sqrt{1+g_1\xi}+\sqrt{1+(\mu+g_2)\xi+\mu g_2\xi^2}} \\ \xi_2^* = \dfrac{\xi\sqrt{1+\xi g_1}}{\sqrt{1+\xi g_1}+\sqrt{1+(\mu+g_2)\xi+\mu g_2\xi^2}} \end{cases}.$$

Thus, the proposition can be concluded. ■

## IV. IMPACTS OF BEAMWIDTH AND SELF-INTERFERENCE CANCELLATION ON MM-WAVE AF RELAYING

Based on the general results presented in Sec. III, in this section, we investigate the impacts of beamwidth and self-interference coefficient, respectively, on the maximum achievable rates of two-hop AF relaying systems with mm-wave.

### A. Impact of Beamwidth

By Propositions 1 and Propositions 2, we know that both $\eta_{\text{HD}}^*$ and $\eta_{\text{FD}}^*$ monotonically increase in $g_i \triangleq |h_i|^2$. It is known from (1) that, $h_i$ is related to the beamwidth of directional antennas, since $G(0)$ and $G(\theta_i)$ are determined by $\theta_m$, as shown in (4). Thus, $\theta_m$ determines $g_i$, and hence affects the rates of relaying systems. It is easy to find that, (4) can be reformulated as

$$G(\theta_i) = G(0)\,\epsilon^{\min\left(1,\left(\frac{2\theta_i}{\theta_m}\right)^2\right)},$$

where $\epsilon = 0.0094$. Thus, $g_i$ can be equivalently written as

$$g_i = \left(\frac{B_i}{1+\tau\theta_m}\right)^2\left|1+z_i\cdot\epsilon^{\min\left(1,\left(\frac{2\theta_i}{\theta_m}\right)^2\right)}\right|^2, \quad (9)$$

where, for notational simplicity, $\tau$, $B_i$ and $z_i$ are respectively given as

$$\begin{cases} \tau \triangleq \dfrac{42.6443}{2\pi} \in \mathbb{R} \\ B_1 \triangleq \dfrac{\lambda}{4\pi\sqrt{(H_S-H_R)^2+L_1^2}} \cdot 10^{2.028} \in \mathbb{R} \\ B_2 \triangleq \dfrac{\lambda}{4\pi\sqrt{(H_R-H_D)^2+L_2^2}} \cdot 10^{2.028} \in \mathbb{R} \\ z_i \triangleq \Gamma(\theta_i) \cdot \cos\theta_i \cdot \exp(-j\Delta\varphi_i) \in \mathbb{C} \end{cases}.$$

In the following lemma, we demonstrate the decaying rule of $g_i$ with respect to $\theta_m$.

**Lemma 1.** *With respect to beamwidth $\theta_m$, channel gain $g_i$ scales as $\mathcal{O}\left(\theta_m^{-2}\right)$.*

*Proof:* Let $\theta_i$ be the given reflection angle. According to (9), we have

$$g_i < \left(\frac{B_i}{\tau}\right)^2 \cdot \left|1 + z_i \cdot \epsilon^{\min\left(1,\left(\frac{2\theta_i}{\theta_m}\right)^2\right)}\right|^2 \cdot \theta_m^{-2}$$

$$\leq \left(\frac{B_i}{\tau}\right)^2 \cdot \left|1 + |z_i| \cdot \epsilon^{\min\left(1,\left(\frac{2\theta_i}{\theta_m}\right)^2\right)}\right|^2 \cdot \theta_m^{-2}$$

$$< \left(\frac{B_i \cdot (1 + |z_i|)}{\tau}\right)^2 \cdot \theta_m^{-2},$$

where the last line is obtained due to the fact that $\min\left(1, \frac{2\theta_i}{\theta_m}\right)$ is a non-increasing function of $\theta_m$. Note that $\epsilon = 0.0094$ and $0 < |z_i| < 1$, it can be concluded that $g_i = \mathcal{O}\left(\theta_m^{-2}\right)$. ∎

Furthermore, we present series of inequalities in the following lemma.

**Lemma 2.** *For $\forall x \geq 0$, we have*

$$\ln(1+x) \leq \frac{x}{\sqrt{1+x}} \leq \min\{\sqrt{x}, x\}.$$

*Proof:* For the first inequality, we define

$$f(x) \triangleq \ln(1+x) - \frac{x}{\sqrt{1+x}},$$

then the first derivative with respect to $x$ is given as

$$\frac{df(x)}{dx} = \frac{2\sqrt{1+x} - x - 2}{2(1+x)^{\frac{3}{2}}} = -\frac{\left(\sqrt{1+x} - 1\right)^2}{2(1+x)^{\frac{3}{2}}}.$$

It is clear that, the first derivative of $f(x)$ over $x \geq 0$ is definitely non-positive, which indicates $f(x) \leq f(0) = 0$ for all $x \geq 0$. For the second inequality, it can be easily concluded by considering if $x \geq 1$ or $0 \leq x \leq 1$. ∎

In light of above two lemmas, the impact of the beamwidth on the maximum achievable rates of the AF relaying systems with mm-wave is revealed in the following theorem.

**Theorem 1.** *With respect to beamwidth $\theta_m$, the maximum achievable rate of two-hop AF relaying with mmWvae (given in Proposition 1 or 2) scales as $\mathcal{O}\left(\min\{\theta_m^{-1}, \theta_m^{-2}\}\right)$.*

*Proof:* Please see Appendix A. ∎

The result in Theorem 1 is unsurprising, and it coincides with the intuition that the directional antenna with a narrower beamwidth enables a higher achievable rate. The main contribution of Theorem 1 is that, we characterize the rates of AF relaying schemes in the concise manner, with respect to the beamwidth of the specific Gaussian-type directional antenna.

Subsequently, we investigate the contribution of the reflection path relative to the LOS path. We know that, $h_i$ consists of the LOS component and the reflection component, i.e.,

$$h_i = \underbrace{\frac{\lambda G(0)}{4\pi L_i}}_{\triangleq h_i^{\text{LOS}} \in \mathbb{R}} + \underbrace{\frac{\lambda G(\theta_i)}{4\pi L_i} \cdot z_i}_{\triangleq h_i^{\text{ref}} \in \mathbb{C}} \in \mathbb{C}.$$

We have $\left|h_i^{\text{LOS}}\right| > \left|h_i^{\text{ref}}\right|$, since the reflection component suffers the reflection loss (partial absorption by reflective material) and the additional path loss (due to extra transmission distance for the reflection path).

We define $g_i^{\text{LOS}} \triangleq \left|h_i^{\text{LOS}}\right|^2$ and $g_i^{\text{ref}} \triangleq \left|h_i^{\text{ref}}\right|^2$. To characterize the contribution of the reflection path relative to the LOS path, we define relative contribution $\zeta_i$ in terms of $g_i^{\text{LOS}}$ and $g_i^{\text{ref}}$ as

$$\zeta_i \triangleq \frac{g_i^{\text{ref}}}{g_i^{\text{LOS}}} \in (0, 1). \tag{10}$$

Note that $\left||h_i| - |h_i^{\text{LOS}}|\right| \leq \left|h_i^{\text{ref}}\right|$ always holds, the lower and upper bounds of $g_i$ in terms of $g_i^{\text{LOS}}$ and $\zeta_i$ are given as

$$\left(1 - \sqrt{\zeta_i}\right)^2 \cdot g_i^{\text{dir}} \leq g_i \leq \left(1 + \sqrt{\zeta_i}\right)^2 \cdot g_i^{\text{dir}}.$$

Given $\epsilon$ and $z_i$, we provide the lower and upper bounds for $\zeta_i$ in the proposition below. Moreover, the non-increasing monotonicity of $\zeta_i$ in $\theta_m$ is also revealed.

**Proposition 3.** *Given reflection angle $\theta_i$, relative contribution $\zeta_i$ defined in (10) is monotonically non-decreasing in $\theta_m$, and we further have*

$$0 < \epsilon^2 |z_i|^2 \leq \zeta_i < |z_i|^2 < 1.$$

*Proof:* By the definition of $\zeta_i$, we have

$$\zeta_i = \frac{G^2(\theta_i)|z_i|^2}{G^2(0)} = \epsilon^{2\min\left(1,\left(\frac{2\theta_i}{\theta_m}\right)^2\right)} \cdot |z_i|^2.$$

Due to the monotonicity $\min\left(1, \frac{2\theta_i}{\theta_m}\right)$ with respect to $\theta_m$, it is straightforward to conclude the bounds for $\zeta_i$, as well as the non-decreasing monotonicity. ∎

Proposition 3 indicates that, ground reflections affects the rate of AF relaying systems more when the directional antenna has a broader beamwidth. It is worth mentioning that, the contribution may be exhibited either in the constructive or the destructive way, since $\zeta_i$ is only counted by the absolute value.

### B. Impact of Self-Interference Cancellation

We know that, the SINR for FD-AF relaying, denoted by $\gamma_{\text{FD}}^*$, heavily relies on $\mu$ (see Proposition 2). In this subsection, we investigate the impact of the self-interference coefficient on the maximum achievable rate of FD-AF mm-wave relaying system, i.e., the impact of $\mu$ on $\eta_{\text{FD}}^* = C(\gamma_{\text{FD}}^*)$.

For simplifying illustration, we start with $\gamma_{\text{FD}}^*$, which can be rewritten as

$$\gamma_{\text{FD}}^* = \frac{A_0}{A_1 + \xi\mu + A_2\sqrt{1+\xi\mu}}, \tag{11}$$

where $A_0$, $A_1$ and $A_2$ respectively denote

$$\begin{cases} A_0 = g_1 g_2 \xi^2 \\ A_1 = 2 + (g_1 + g_2)\xi \\ A_2 = 2\sqrt{(1+g_1\xi)(1+g_2\xi)} \end{cases}.$$

In the following lemma, we demonstrate the convexity and monotonicity of $\gamma_{\text{FD}}^*$.

**Lemma 3.** *SINR $\gamma_{\text{FD}}^*$ is strictly convex and monotonically decreasing in self-interference coefficient $\mu \in (0,1)$.*

*Proof:* In terms of $\mu$, we can verify that, $\gamma_{\text{FD}}^*$ is continuous and twice differentiable over $\mu \in (0,1)$. The first-order and second-order partial derivatives of (11), with respect to $\mu$, are respectively given as

$$\frac{\partial \gamma_{\text{FD}}^*}{\partial \mu} = -\frac{\xi A_0 \left(1 + \frac{A_2}{2\sqrt{1+\xi\mu}}\right)}{\left(A_1 + \xi\mu + A_2\sqrt{1+\xi\mu}\right)^2}$$

and

$$\frac{\partial^2 \gamma_{\text{FD}}^*}{\partial \mu^2} \triangleq \frac{\partial}{\partial \mu}\left(\frac{\partial \gamma_{\text{FD}}^*}{\partial \mu}\right)$$
$$= \frac{\xi^2 A_0 \left(3A_2^2 \sqrt{1+\xi\mu} + 8(1+\xi\mu)^{\frac{3}{2}} + A_2(8+A_1+9\xi\mu)\right)}{4(1+\xi\mu)^{\frac{3}{2}} \left(A_1 + \xi\mu + A_2\sqrt{1+\xi\mu}\right)^3}.$$

Evidently, we have

$$\frac{\partial \gamma_{\text{FD}}^*}{\partial \mu} < 0 \text{ and } \frac{\partial^2 \gamma_{\text{FD}}^*}{\partial \mu^2} > 0,$$

which concludes the decreasing monotonicity and convexity, respectively. ∎

The decreasing monotonicity of $\gamma_{\text{FD}}^*$ in $\mu$, revealed in Lemma 3, indicates the an upper bound for SINR, i.e.,

$$\gamma_{\text{FD}}^* < \overline{\gamma}_{\text{FD}}^* \triangleq \lim_{\mu \to 0} \gamma_{\text{FD}}^* = \frac{A_0}{A_1 + A_2}.$$

Given any self-interference coefficient $\mu > 0$, to characterize the difference between $\gamma_{\text{FD}}^*$ and $\overline{\gamma}_{\text{FD}}^*$, we define the ratio of $\gamma_{\text{FD}}^*$ to $\overline{\gamma}_{\text{FD}}^*$, i.e.,

$$\kappa \triangleq \frac{\overline{\gamma}_{\text{FD}}^*}{\gamma_{\text{FD}}^*} = \frac{A_1 + \xi\mu + A_2\sqrt{1+\xi\mu}}{A_1 + A_2}, \quad (12)$$

and the increasing monotonicity of $\kappa$ with respect to sum-power constraint $\xi$ is presented in Proposition 4.

**Proposition 4.** *Given channel gain $g_i$ on the $i^{\text{th}}$ hop and any $\mu > 0$, the difference metric $\kappa$ defined in (12) is monotonically increasing with respect to sum-power constraint $\xi$.*

*Proof:* Please see Appendix B. ∎

Proposition 4 indicates that, in the presence of imperfect self-interference cancellation, i.e., $\mu \neq 0$, the performance difference between $\gamma_{\text{FD}}^*$ and $\overline{\gamma}_{\text{FD}}^*$, characterized by $\kappa$ in the format of ratio, gets enlarged when increasing the sum-power budget. Therefore, FD-AF relaying with a smaller $\mu$ needs to be devised especially when sum-power budget $\xi$ is high, such that all the power can be fully exploited for further improving the achievable rate.

Based on Lemma 3, the monotonicity and the convexity of $\eta_{\text{FD}}^*$ with respect to $\mu$ are shown in the following theorem.

**Theorem 2.** *Achievable rate $\eta_{\text{FD}}^*$ by Proposition 2 is strictly convex and monotonically decreasing in self-interference coefficient $\mu \in (0,1)$.*

*Proof:* Please see Appendix C. ∎

The convexity and monotonicity revealed in Theorem 2 indicate the great importance of reducing the self-interference coefficient for FD-AF mm-wave relaying.

Besides, it is revealed that $\eta_{\text{FD}}^* = \mathcal{O}\left(\mu^{-\frac{1}{2}}\right)$ in Theorem 3.

**Theorem 3.** *Achievable rate $\eta_{\text{FD}}^*$ scales as $\mathcal{O}\left(\mu^{-\frac{1}{2}}\right)$, with respect to $\mu \in (0,1)$.*

*Proof:* Based on (11), we have

$$\eta_{\text{FD}}^* = C\left(\frac{A_0}{A_1 + \xi\mu + A_2\sqrt{1+\xi\mu}}\right)$$
$$< C\left(\frac{A_0}{A_1 + \xi\mu + A_2\sqrt{\xi\mu}}\right).$$

It is worth noting that, $\sqrt{\mu} > \mu$ can be achieved for $\mu \in (0,1)$. Applying Lemma 2, we further obtain that

$$\eta_{\text{FD}}^* < C\left(\frac{A_0}{\xi + A_2\sqrt{\xi}} \cdot \mu^{-1}\right) \leq \frac{1}{\ln 2}\sqrt{\frac{A_0}{\xi + A_2\sqrt{\xi}}} \cdot \mu^{-\frac{1}{2}}.$$

Thus, we can conclude that $\eta_{\text{FD}}^* = \mathcal{O}\left(\mu^{-\frac{1}{2}}\right)$. ∎

It can be seen from Theorem 3 that, the maximum achievable rate of FD-AF mm-wave relaying acceleratingly increases as reducing the self-interference coefficient, which is in accordance with the result by Theorem 2, and again demonstrates the huge benefit of suppressing the self-interference.

The following proposition gives the condition of $\mu$ for FD-AF to outperform direct transmission or HD-AF relaying.

**Proposition 5.** *Let $\chi \geq 0$ be the given maximum achievable rate of direct transmission or HD-AF relaying. For $\eta_{\text{FD}}^* \leq \chi$, we have*

$$\mu \in \begin{cases} (0,1), & \Psi_\chi \leq 1 \\ \left[\frac{\Psi_\chi^2 - 1}{\xi}, 1\right), & 1 < \Psi_\chi < \sqrt{1+\xi} \\ \emptyset, & \Psi_\chi \geq \sqrt{1+\xi} \end{cases}$$

*where $\Psi_\chi$ is given as*

$$\Psi_\chi \triangleq \xi\sqrt{g_1 g_2 \left(1 - \frac{1}{2^\chi - 1}\right)} - \sqrt{(1+g_1\xi)(1+g_2\xi)}.$$

*Proof:* For $\eta_{\text{FD}}^* = C(\gamma_{\text{FD}}^*) \leq \chi$, we have

$$\gamma_{\text{FD}}^* = \frac{A_0}{A_1 + \xi\mu + A_2\sqrt{1+\xi\mu}} \leq 2^\chi - 1.$$

Applying the change of variables $v = \sqrt{1+\xi\mu}$, it is easy

TABLE I
SYSTEM SETTINGS

| Parameter | Notation |
|---|---|
| Wavelength | $\lambda = 5 \times 10^{-3}$ m |
| Sum-Power Constraint | $\xi$ |
| Main-lobe Beamwidth | $\theta_m$ |
| Self-interference Coefficient | $\mu$ |
| Deployment Height | $H = 5$ m |
| Source-Destination Distance | $L = 200$ m |
| Source-Relay Distance | $L_1$ |
| Ground Dielectric Constant | $\omega \approx 15$ [35] |

to obtain that

$$v \geq \sqrt{1 + \frac{A_0}{2^{\chi} - 1} - A_1 + \frac{A_2^2}{4}} - \frac{A_2}{2}$$
$$= \xi\sqrt{g_1 g_2 \left(1 - \frac{1}{2^{\chi} - 1}\right)} - \sqrt{(1 + g_1\xi)(1 + g_2\xi)} \triangleq \Psi_{\chi}.$$

According to the possible values of $\Psi_{\chi}$ and the condition that $1 < v \leq \sqrt{1+\xi}$, we can solve $\mu$ via recovering $\mu$ in terms of $v$, i.e., $\mu = \xi^{-1}(v^2 - 1)$, which then completes the proof of the proposition. ∎

Briefly, given channel gains $g_i$ ($i = 1, 2$) and sum-power constraint $\xi$, $\Psi_{\chi}$ characterizes the resulting performance relative to threshold $\chi$. Regarding the range of $\Psi_{\chi}$, critical value 1 characterizes the minimum requirement on channel gains and sum power for $\Psi_{\chi}$ to achieve $\chi$ (given perfect self-interference cancellation, i.e., $\mu \to 0$), while critical value $\sqrt{1+\xi}$ characterizes the maximum requirement on channel gains and sum power for $\Psi_{\chi}$ to achieve $\chi$ (assuming no self-interference cancellation, i.e., $\mu \to 1$). Thus, by Proposition 5, $\eta_{\text{FD}}^*$ is definitely less than $\chi$ if channel gains and sum power for $\Psi_{\chi}$ cannot meet the minimum requirement, i.e., $\Psi_{\chi} \leq 1$, while $\eta_{\text{FD}}^*$ definitely exceeds $\chi$ if channel gains and sum power for $\Psi_{\chi}$ can meet the maximum requirement, i.e., $\Psi_{\chi} \geq \sqrt{1+\xi}$. For $\Psi_{\chi}$ between 1 and $\sqrt{1+\xi}$, $\eta_{\text{FD}}^*$ can achieve the given $\chi$, conditioning on certain feasible set for $\mu$, i.e., $\xi^{-1}(\Psi_{\chi}^2 - 1) \leq \mu < 1$.

## V. NUMERICAL RESULTS

In this section, the maximum achievable rates of two AF relaying schemes with mm-wave by Proposition 1 and 2, respectively, are demonstrated. Moreover, the impacts of the beamwidth and the self-interference coefficient are shown. We assume the used mm-wave radio is perpendicularly polarized (refer to (2)). In addition, we assume that nodes $S$, $R$ and $D$ are placed in a line, i.e., $L_1 + L_2 = L$, and deployed at with the identical height, i.e., $H_S = H_R = H_D = H$. Several key system parameters and corresponding notations are summarized in Table I.

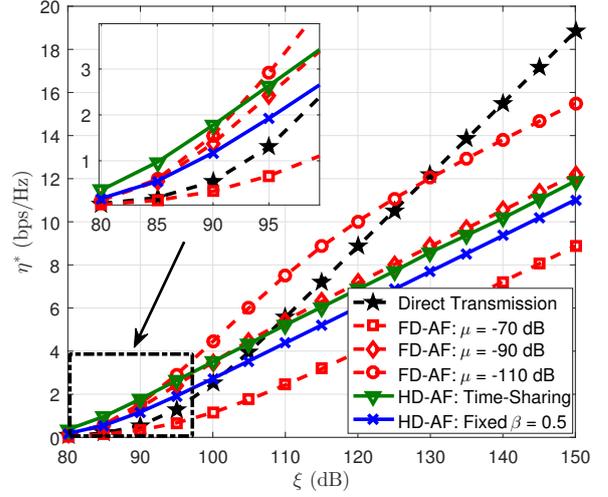

Fig. 2. Maximum rates of direct transmission and both AF relaying schemes under different sum-power constraints, where $L_1 = 80$ m and $\theta_m = \frac{\pi}{6}$.

### A. Performance Comparison

With respect to sum-power constraint $\xi$, the maximum achievable rates of direct transmission, HD-AF and FD-AF mm-wave relaying systems are provided in Fig. 2. For HD-AF relaying, rate $\eta_{\text{HD}}^*$ by Proposition 1 and that with fixed $\beta = \frac{1}{2}$ (refer to (6)) are both provided. It is shown that, compared to the case of fixed $\beta = \frac{1}{2}$, a performance gain, i.e., $\approx 5$ dB, can be achieved when applying the optimized time-sharing scheme. For FD-AF relaying, according to Proposition 2, the achievable rates with different self-interference coefficients are shown. Evidently, varying $\mu$ from $-70$ dB to $-110$ dB, we find that $\eta_{\text{FD}}^*$ can be significantly improved when $\mu$ reduces. Comparing the cases with different values of $\mu$, it can be seen that, the self-interference coefficient plays a crucial role in the rate of FD-AF relaying with mm-wave, particularly when a medium or high sum-power budget is given, i.e., $\xi \geq 100$ dB, while the impacts of the self-interference coefficient are relative smaller for $\xi \leq 90$ dB. To be more precise, taking the performance of HD-AF relaying as a benchmark, for $\xi \geq 100$ dB, we notice that $\eta_{\text{FD}}^*$ is much less than $\eta_{\text{HD}}^*$ when $\mu = -70$ dB, while it evidently outperforms the latter when $\mu = -90$ dB or smaller. However, within the low $\xi$ region, i.e., $\xi \leq 90$ dB, it can be seen that, $\eta_{\text{FD}}^*$ is still inferior to $\eta_{\text{HD}}^*$ even though $\mu$ has been reduced to $-110$ dB. When $\xi$ is lower, e.g., $\xi \leq 100$ dB, direct transmission is inferior to HD-AF but beyond FD-AF with $\mu = -70$ dB, while it surpasses the relaying counterparts if $\xi$ is sufficiently high, e.g., $\xi \geq 130$ dB. The findings above indicate that, given high sum-power budget, direct transmission is the best choice. For relaying schemes, FD-AF relaying with a small self-interference coefficient should be adopted for scenarios with medium sum-power budget, while HD-AF relaying is more suitable for scenarios with low sum-power budget. Intuitively, HD-AF largely benefits from the absence of interference at the low sum-power region, and FD-AF with a small self-interference coefficient largely benefits from the full utilization of time slots for transmission at the intermediate sum-power region. However, once the sum-power is sufficiently high, the severe path loss from the source to the destination can be completely compensated, such that direct transmission in mm-wave communications becomes the most efficient option, instead of using any relaying technique.

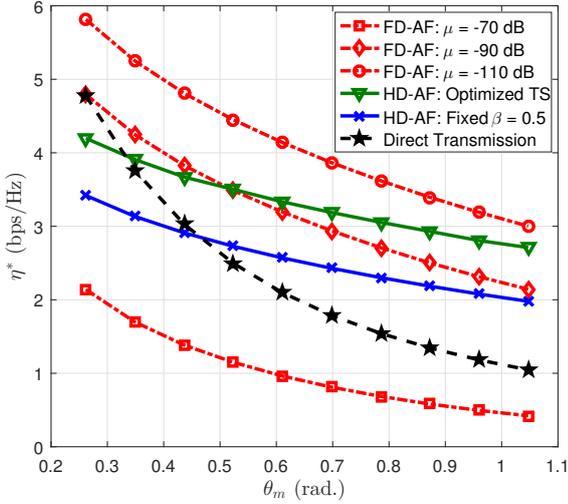

Fig. 3. Maximum rates of AF relaying schemes (FD-AF, HD-AF with time-sharing, and HD-AF with $\beta = \frac{1}{2}$) with different beamwidth under a sum-power constraint $\xi = 100$ dB are applied, and $L_1 = 80$ m.

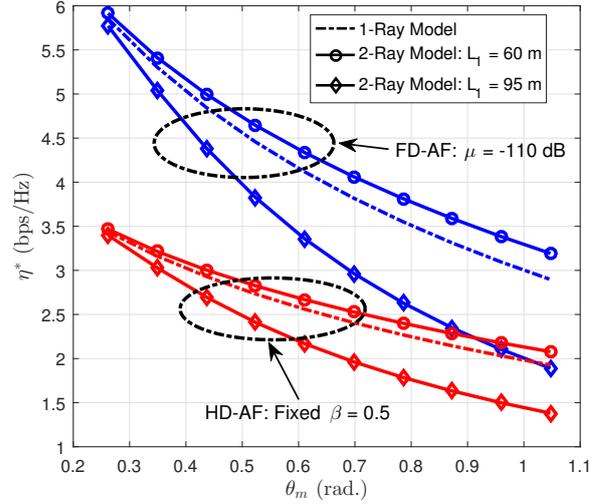

Fig. 4. Contribution of ground reflections to rates with respect to different beamwidth, where $\xi = 100$ dB.

### B. Impacts of Beamwidth

The impacts of beamwidth on the rates of different transmission schemes are shown in Fig. 3, where the direct transmission scheme and both AF relaying schemes are compared. Clearly, all schemes suffer performance degradation as $\theta_m$ increases, while the direct transmission encounters the most severe deterioration, and FD-AF follows. It can be seen that, for both FD-AF and HD-AF relaying schemes, the rates degrades with $\theta_m$ as $\mathcal{O}\left(\min\left\{\theta_m^{-1}, \theta_m^{-2}\right\}\right)$, as analyzed in Theorem 1. In addition, compared to $\eta_{\text{HD}}^*$, we notice that $\eta_{\text{FD}}^*$ rapidly decays when $\theta_m$ grows, which indicates the performance by FD-AF mm-wave relaying is more sensitive to the variation of beamwidth. The slow decay of $\eta_{\text{HD}}^*$ with respect to $\theta_m$ mainly comes from the time fractions in fulfilling the two-phase transmission (in contrast to the full utilization of time duration in $\eta_{\text{FD}}^*$), which mitigate the impact of the variation of $g_1$ and/or $g_2$ on the achievable rate. Furthermore, higher achievable rates can be achieved by FD-AF mm-wave relaying, when a smaller self-interference coefficient (e.g., $\mu \le -90$ dB) and a narrower beamwidth of directional antenna (e.g., $\theta_m \le \frac{\pi}{6}$) are provided. Otherwise, HD-AF mm-wave relaying or direct transmission strategy should be adopted. We can also see that, given $\xi = 100$ dB, when applying very sharp beams, i.e., small $\theta_m$, direct transmission outperforms the HD-AF relaying. Note that, narrowing the beamwidth increases the channel gain, which is equivalent to elevating the sum-power budget while keeping the channel gain untouched. In this sense, the gain by sharpening beams is equivalent to the gain by increasing the sum-power budget with fixing the channel gains. Recalling that direct transmission outperforms HD-AF in the high sum-power region (as shown in Fig. 2), the observation that HD-AF is inferior to direct transmission when applying narrower beams then can be explained. Therefore, the findings above demonstrate that, to achieve higher rates of FD-AF mm-wave relaying, a careful joint treatment of the beamwidth and the self-interference coefficient during system designs and implementations is important. Besides, direct transmission is still a promising candidate when the antennas are highly directional.

Fig. 4 illustrates the contribution of ground reflections in two AF relaying schemes with mm-wave. For comparison, we take the curves labeled with "1-ray" as references, which correspond to the conventional modeling method that considers the LOS path only for mm-wave channels. Evidently, with the specific two-ray model (labeled with "2-ray"), the rates heavily rely on $L_1$ and $L_2$. More precisely, given $\theta_m$, for $L_1 = 60$ m (simultaneously, $L_2 = L - L_1 = 140$ m), the achieved rate is beyond that under the "1-ray" model. However, the rates is remarkably inferior to that under the "1-ray" model when $L_1 = 95$ m (simultaneously, $L_2 = 105$ m). This surprisingly significant performance fluctuation stems from the contribution of ground reflections. It can be seen from (1) that, the reflection component of the channel coefficient depends on the incident angle of reflected signal, the radiation pattern of directional antenna, and the transmission distance. Thus, for $L_1 = 60$ m, the ground reflections contribute to the channel gain in the constructive manner, thereby improving the rates of the two-hop mm-wave relaying system, while the ground reflections for $L_1 = 95$ m adversely affect the rate. In light of above findings, an important insight for practical applications here is that, when the beamwidth is not very small, considerable improvement can be achieved via properly deploying the relay, such that the constructive ground reflections can be fully exploited. Furthermore, we can see that, when $\theta_m$ decreases, the rates with the two-ray model converges to that with the "1-ray" model. This observation indicates that, the impacts of ground reflections diminish when highly directional antennas are employed, and it agrees to Proposition 3. Hence, from Fig. 4, it can be concluded that, the two-ray model provides a general framework for arbitrary beamwidth, while the "1-ray" model is only suitable for scenarios with very narrow beams.

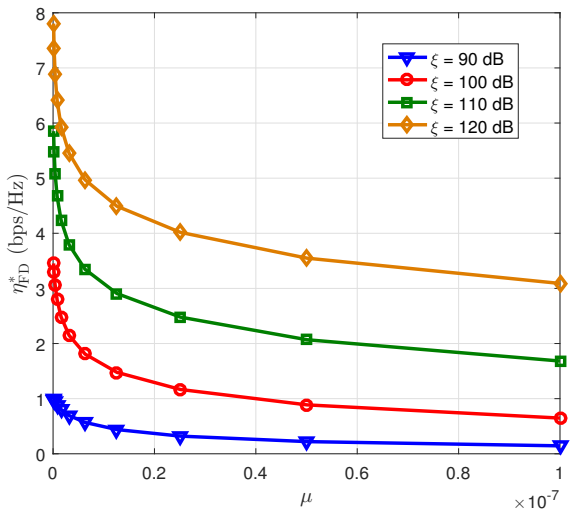

Fig. 5. Impact of $\mu$ on $\eta_{\text{FD}}^*$, where $\theta_m = \frac{\pi}{4}$ and $L_1 = 100$ m.

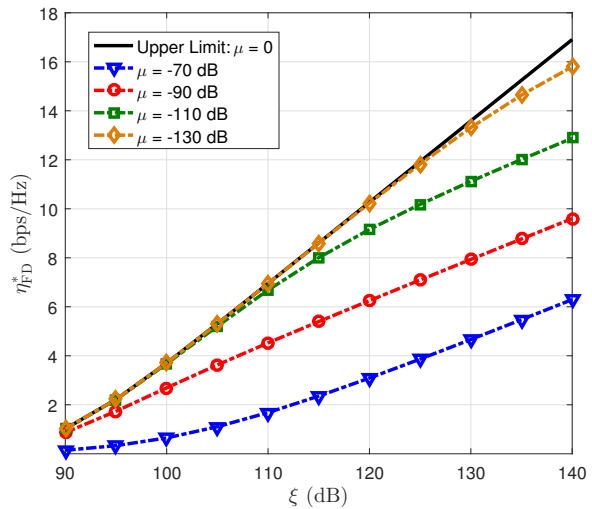

Fig. 6. Performance gap between $\eta_{\text{FD}}^*$ and its upper limit $\overline{\eta}_{\text{FD}}^*|_{\mu=0}$, where $\theta_m = \frac{\pi}{4}$ and $L_1 = 100$ m.

### C. Impacts of Self-Interference Cancellation

Fig. 5 depicts the impact of the self-interference coefficient on the rate of FD-AF mm-wave relaying. We find that, $\eta_{\text{FD}}^*$ dramatically drops when increasing $\mu$ from $-120$ dB to $-70$ dB. As aforementioned in Theorem 2, $\eta_{\text{FD}}^*$ with respect to $\mu$ varies in the convex and decreasing manner. Thus, a slight reduction of $\mu$ is capable of providing remarkable performance improvement, and the improvement accelerates at lower $\mu$ and higher $\xi$, as shown in Fig. 5. This finding indicates the great importance of devising powerful self-interference cancellation techniques for FD-AF mm-wave relaying, especially when the two-hop AF mm-wave relaying system is assigned with a higher sum-power budget.

With respect to different sum-power constraints, in Fig. 6, we compare the performance of FD-AF mm-wave relaying in the presence of non-zero $\mu$, and its upper limit is achieved by assuming $\mu = 0$. Given sum-power constraint $\xi$, when $\mu$ decreases from $-70$ dB to $-130$ dB, it is not surprising that $\eta_{\text{FD}}^*$ gradually approaches its upper limit $\overline{\eta}_{\text{FD}}^*$. We notice that, with any given $\mu$, the rate difference between $\eta_{\text{FD}}^*$ and $\overline{\eta}_{\text{FD}}^*$ (vertical performance gap) gets enlarged as $\xi$ grows, which is in accordance with Proposition 4. Moreover, in the presence of larger self-interference coefficients, it is evident that only a smaller $\xi$ is required for $\eta_{\text{FD}}^*$ to approach its upper limit $\overline{\eta}_{\text{FD}}^*$. This interesting observation can be explained by (11), since the SINR approaches its upper limit only if the product term $\mu\xi$ is sufficiently small. Therefore, a much smaller $\mu$ will be accordingly required for any given higher $\xi$, such that the condition $\mu\xi \to 0$ can be satisfied to make $\eta_{\text{FD}}^*$ approach $\overline{\eta}_{\text{FD}}^*$. In light of the relation between $\mu$ and $\xi$ for guaranteeing $\mu\xi \to 0$, an important insight for mm-wave relay implementation is that, when the given sum-power budget is not high, it is not necessary to devise strong self-interference cancellation techniques, since the limiting performance can be easily approached even if $\mu$ is not sufficiently small. Thereby, the implementation cost and complexity of FD-AF relays can be largely reduced.

## VI. CONCLUSIONS

We have studied the achievable rate of the two-hop AF relaying system with mm-wave, where two relaying schemes, i.e., HD-AF and FD-AF relaying are considered. With the joint treatment of the two-ray mm-wave channel and Gaussian-type directional antenna, we have investigated the impact of the beamwidth and the self-interference coefficient on the maximum achievable rate. Under a sum-power constraint, it has been demonstrated that, FD-AF relaying outperforms its HD counterpart, only when directional antennas with a smaller beamwidth and the AF relay with a smaller self-interference coefficient are applied. Thus, HD-AF relaying scheme is still a competitive candidate for mm-wave communications whenever above conditions cannot be satisfied. We also have found that, when the sum-power budget is sufficiently high or the beam of directional antenna is sufficiently sharp, the performance of direct transmission is beyond that of two relaying schemes, such that AF relaying is not necessary for in this case. In addition, it has been demonstrated that, for both mm-wave relaying schemes, the maximum achievable rate scales as $\mathcal{O}\left(\min\left\{\theta_m^{-1}, \theta_m^{-2}\right\}\right)$ with respect to beamwidth $\theta_m$, and scales as $\mathcal{O}\left(\mu^{-\frac{1}{2}}\right)$ for FD-AF relaying with respect to self-interference coefficient $\mu$. Furthermore, it has been revealed that, the ground reflection may significantly affect the performance of mm-wave communications, constructively or destructively. Therefore, it is crucial to incorporate the impacts of ground reflections for analyzing or designing mm-wave networks with directional antennas.

## APPENDIX A
## PROOF OF THEOREM 1

By Lemma 1, it is known that, there exists some positive $K_i$, such that for all feasible $\theta_m$ we have

$$g_i \leq K_i \cdot \theta_m^{-2}.$$

For HD-AF relaying, we know that $\eta_{\text{HD}}^*$ is monotonically increasing with $g_i$. By Proposition 1, we obtain that

$$\eta_{\text{HD}}^* \stackrel{(a)}{<} \max_{\substack{0 \leq \beta \leq 1 \\ \xi_1, \xi_2 \geq 0 \\ \beta \xi_1 + \bar{\beta} \xi_2 = \xi}} \min \{\beta C(g_1 \xi_1), (1-\beta) C(g_1 \xi_1)\}$$

$$= \max_{\substack{0 \leq \beta \leq 1, \xi_1 \geq 0 \\ \beta \xi_1 \leq \xi}} \min \{\beta C(g_1 \xi_1), (1-\beta) C(g_1 \xi_1)\}$$

$$\stackrel{(b)}{\leq} \max_{\substack{0 \leq \beta \leq 1, \xi_1 \geq 0 \\ \beta \xi_1 \leq \xi}} \beta C(g_1 \xi_1),$$

where $(a)$ applies the fact that

$$\frac{g_1 g_2 \xi_1 \xi_2}{1 + g_1 \xi_1 + g_2 \xi_2} = \frac{g_2 \xi_2}{1 + g_1 \xi_1 + g_2 \xi_2} \cdot g_1 \xi_1 < g_1 \xi_1,$$

and $(b)$ follows from the property that $\min\{a, b\} \leq a$ for any $a, b \in \mathbb{R}$. By Lemma 2, it is easy to obtain that

$$C(g_1 \xi_1) < \log_2 \left(1 + K_1 \xi_1 \theta_m^{-2}\right)$$
$$\leq \frac{1}{\ln 2} \cdot \min \left\{K_1 \xi_1 \theta_m^{-2}, \sqrt{K_1 \xi_1} \theta_m^{-1}\right\},$$

which subsequently yields

$$\eta_{\text{HD}}^* < \max_{\substack{0 \leq \beta \leq 1, \xi_1 \geq 0 \\ \beta \xi_1 \leq \xi}} \frac{K_1 \beta \xi_1}{\ln 2} \theta_m^{-2} = \frac{K_1 \xi}{\ln 2} \cdot \theta_m^{-2} = \mathcal{O}\left(\theta_m^{-2}\right).$$

or

$$\eta_{\text{HD}}^* < \max_{\substack{0 \leq \beta \leq 1, \xi_1 \geq 0 \\ \beta \xi_1 \leq \xi}} \frac{\beta \sqrt{K_1 \xi_1}}{\ln 2} \theta_m^{-1} \leq \frac{\sqrt{K_1 \xi}}{\ln 2} \cdot \theta_m^{-1} = \mathcal{O}\left(\theta_m^{-1}\right).$$

Since $\mathcal{O}\left(\theta_m^{-2}\right) \subset \mathcal{O}\left(\theta_m^{-1}\right)$ as $\theta_m \geq 1$, we then have $\eta_{\text{HD}}^* = \mathcal{O}\left(\theta_m^{-1}\right)$ for all $\theta_m > 0$, or $\eta_{\text{HD}}^* = \mathcal{O}\left(\theta_m^{-2}\right)$ particularly when $\theta_m \geq 1$.

Likewise, for FD-AF relaying, we have

$$\eta_{\text{FD}}^* < C\left(\frac{\xi K_1 K_2 \cdot \theta_m^{-2}}{K_1 + K_2 + 2\sqrt{K_1 K_2 (1 + \mu \xi)}}\right)$$
$$\leq \frac{1}{\ln 2} \sqrt{\frac{\xi K_1 K_2}{K_1 + K_2 + 2\sqrt{K_1 K_2 (1 + \mu \xi)}}} \cdot \theta_m^{-1} = \mathcal{O}\left(\theta_m^{-1}\right).$$

Again, it is easy to see that, $\eta_{\text{FD}}^* \in \mathcal{O}\left(\theta_m^{-2}\right)$ if $\theta_m \geq 1$.

## APPENDIX B
## PROOF OF PROPOSITION 4

Note that $\kappa$ can be reformulated as

$$\kappa = 1 + \underbrace{\frac{\xi \mu}{A_1 + A_2}}_{\triangleq \kappa_1} + \underbrace{\frac{A_2\left(\sqrt{1 + \xi \mu} - 1\right)}{A_1 + A_2}}_{\triangleq \kappa_2},$$

where $\kappa_1$ and $\kappa_2$ are exactly shown as

$$\kappa_1 = \frac{\xi \mu}{\left(\sqrt{1 + g_1 \xi} + \sqrt{1 + g_2 \xi}\right)^2}$$

and

$$\kappa_2 = \frac{2\sqrt{(1 + g_1 \xi)(1 + g_2 \xi)} \cdot \left(\sqrt{1 + \xi \mu} - 1\right)}{\left(\sqrt{1 + g_1 \xi} + \sqrt{1 + g_2 \xi}\right)^2}.$$

For the first-order partial derivatives of $\kappa_1$ and $\kappa_2$ with respect to $\xi$, it is easy to verify that

$$\frac{\partial \kappa_1}{\partial \xi} > 0 \text{ and } \frac{\partial \kappa_2}{\partial \xi} > 0,$$

which naturally yields

$$\frac{\partial \kappa}{\partial \xi} = \frac{\partial \kappa_2}{\partial \xi} + \frac{\partial \kappa_2}{\partial \xi} > 0.$$

Therefore, $\kappa$ is a monotonically increasing function of $\xi$, which concludes the proposition.

## APPENDIX C
## PROOF OF THEOREM 2

For $\eta_{\text{FD}}^*$, its first-order partial derivative is

$$\frac{\partial \eta_{\text{FD}}^*}{\partial \mu} = \frac{\partial \eta_{\text{FD}}^*}{\partial \gamma_{\text{FD}}^*} \cdot \frac{\partial \gamma_{\text{FD}}^*}{\partial \mu} = \frac{1}{(1 + \gamma_{\text{FD}}^*) \ln 2} \cdot \frac{\partial \gamma_{\text{FD}}^*}{\partial \mu}.$$

Likewise, the second-order partial derivative is given by

$$\frac{\partial^2 \eta_{\text{FD}}^*}{\partial \mu^2} = \frac{\partial}{\partial \mu}\left(\frac{\partial \eta_{\text{FD}}^*}{\partial \gamma_{\text{FD}}^*} \cdot \frac{\partial \gamma_{\text{FD}}^*}{\partial \mu}\right)$$
$$= \frac{\partial^2 \eta_{\text{FD}}^*}{\partial \gamma_{\text{FD}}^{*2}} \cdot \left(\frac{\partial \gamma_{\text{FD}}^*}{d\mu}\right)^2 + \frac{\partial \eta_{\text{FD}}^*}{\partial \gamma_{\text{FD}}^*} \cdot \frac{\partial^2 \gamma_{\text{FD}}^*}{\partial \mu^2}$$
$$= \frac{1}{(1 + \gamma_{\text{FD}}^*) \ln 2} \cdot \frac{\partial^2 \gamma_{\text{FD}}^*}{\partial \mu^2} - \frac{1}{(1 + \gamma_{\text{FD}}^*)^2 \ln 2} \cdot \left(\frac{\partial \gamma_{\text{FD}}^*}{\partial \mu}\right)^2,$$

where the chain rule of derivatives for composite functions is applied. With the aid of derivations in Lemma 3, it is not difficult to verify that,

$$\frac{\partial \eta_{\text{FD}}^*}{\partial \mu} < 0 \text{ and } \frac{\partial^2 \eta_{\text{FD}}^*}{\partial \mu^2} > 0.$$

Therefore, the decreasing monotonicity and the strict convexity are concluded, respectively.